# How can one use a two-component Bose-Einstein condensates to operationally ‘bypass’ the no-cloning theorem?

**Shouvik Datta***

*Department of Physics, Indian Institute of Science Education and Research - Pune,*

*Pune 411008, Maharashtra, India*

*Email: shouvik@iiserpune.ac.in

**Abstract**

The no-cloning theorem in quantum cryptography prevents any eavesdropper from exactly duplicating an arbitrary quantum superposition state of a single photon. Here we argue that an experimental scheme to produce an interacting, two-component Bose-Einstein condensates can, in principle, generate macroscopically large number of bosonic clones of any arbitrary single photon wave packet with high fidelity at large N limit of thermodynamic equilibrium using (say) excitons or electron-hole pairs. It is possible, because initially one can isolate the two orthogonal polarizations using polarizing beam splitters and then amplify the corresponding single photon wave packets identically but separately. This is to ensure that the amplified beams can be used to generate proportionately same, yet large numbers of bosons (say, excitons) to produce two distinct but interacting condensates using additional light-matter interactions within a semiconductor structure. One can then extract the cloned photons once the identical excitons in the two-component quantum ground state of the condensate radiatively recombine. Thus the overall cloning process can operationally ‘bypass’ the restrictions imposed by the above-mentioned theorem. This is because the quantum statistical nature of this proposed ‘cloning operation’ does not require any strict unitary evolution of standard quantum mechanics within a single Hilbert space.

## 1. Introduction

The original no-cloning theorem [1,2] states that linearity and unitarity of quantum mechanics prevents exact copying of an arbitrary quantum superposition state. As such various replicating schemes were proposed and reviewed [3] for cloning 'albeit imperfectly' [4,5] up to an optimal level. However, most reports of 'optimal' cloning of photons in quantum cryptography relied heavily on the process of duplication *'only and only'* inside the constraints of norm preserving unitary evolution within a single, 1st quantized Hilbert space of quantum mechanics. This is, to the best of our understanding, need not be followed exactly from an 'operational' perspective to engineer 'quantum cloning' involving a series of intermediate quantum statistical mechanical processes. We will also argue that the expected fidelity of the final cloned quantum state with respect to the initial quantum state of a single photon can, however, be preserved by ensuring 'identical' but separate amplifications of both quantum channels (i.e. two orthogonal polarizations) and subsequently generating quantum superposition of a two-component interacting Bose-Einstein condensates (BEC) having a macroscopically large number of identical bosonic clones in its ground state. To elaborate further, here we first take the example of forming a quantum superposition state of two, interacting atomic BECs [6]. We then extend that to generate BEC of quasi atomic entity like excitons or bound electron-hole pairs to execute such quantum cloning using semiconductor structures. However, the proposed experimental scheme can also be implemented using BECs of other physical systems including those with polaritons, atoms, photons etc. In fact, we will show that generation of such interacting, two component, quantum superposition states of macroscopically large number of excitons as tensor product of single-particle states are well documented for condensates of Helium-3 and Rubidium atoms in particular.

In this experimental framework, one will not be violating any linearity, unitarity of 1st quantized formalism of norm-preserving Hilbert space while executing this quantum cloning by creating two interacting excitonic BECs. Formation of any BEC is essentially outside the domain of single-particle quantum mechanics and primarily governed by the many-body physics of quantum statistical mechanics. It is mainly because of this, the proposed 'quantum copier' which processes this quantum cloning mechanism can be 'operationally' mapped from the 1st quantized state of the incoming single photon to a set of 2$^{nd}$ quantized Fock spaces or many-body Hilbert spaces of an interacting, two-component BEC of macroscopically large number of bosons using light-matter interactions. In our case, these bosons are large (say, N) numbers of excitons in the quantum ground state of that two-component BEC. So, here in this framework, creations and destructions of particles (e.g. excitons) are allowed through a sequence of transitions and these do not remain within the scope of unitary evolutions. Operationally speaking, in this experimental scheme, a single photon input in an arbitrary quantum superposition state $|\Psi\rangle$ is first isolated into its independent polarization states using polarising beam splitters. Subsequently, '*identical*' but separate and large amplifications of these two individual polarization states are executed without running in to any conflict with quantum mechanics. After amplifications, these two types of photons are used to create two distinct sets of proportionately large numbers of excitons within a semiconducting structure. Subsequently these two sets of 'macroscopically large' numbers of excitons are driven towards the quantum mechanical ground state of a two-component, interacting BEC via 'weak' Josephson like light-matter interactions as prescribed [6] in the past for atomic BECs. Finally, the generic quantum clones of excitons in that quantum ground state of that two-component BEC can radiatively recombine to germinate multiple copies of the incoming photon. Most importantly, this quantum statistical nature of creating a two-component interacting BEC can allow one to execute the 'operational' processes of this experimental scheme by stepping outside the stringent theoretical constraint of 'unitary' evolutions of quantum mechanics within a single Hilbert space in the first place. Just to emphasize once more, this particular constraint of unitarity and linearity of quantum mechanics was the all-important foundation for the well-known no-cloning theorem [1,2]. Most theoretical and experimental efforts in quantum communication, however, remained heavily restricted by this in the past. Therefore, we will argue why this experimental scheme using BEC as a mediator for quantum cloning can effectively circumvent these restrictions and thereby bypass the no-cloning theorem using a semiconductor based excitonic material/device.

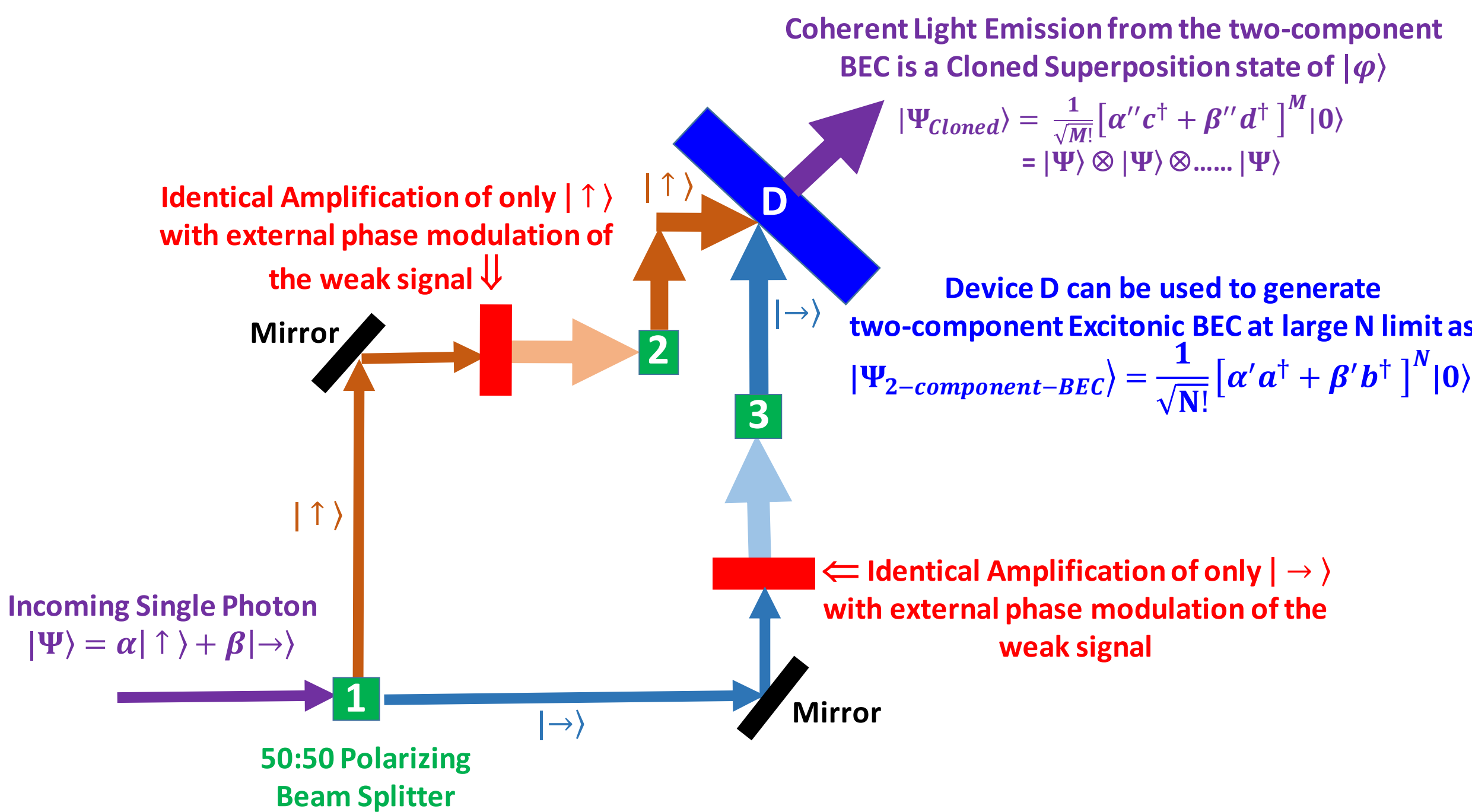

**Figure 1.** Schematic of the quantum cloning process using 'bosonic clones' in the quantum ground state of a two-component, interacting BEC of (say) excitons. We purposely desist from knowing/tracking any time resolved individual outcome(s) of the beam splitting, reflections from the mirrors as well as that of the amplification processes. We treat the whole framework as a black box till the quantum ground state of the two-component, interacting BECs emit those cloned photons in a phase coherent way at the large N limit of thermodynamic equilibrium. This is necessary so that we do not to disturb the phase coherence between the vertical $|\uparrow\rangle$ and horizontal $|\rightarrow\rangle$ polarizations by the process of measurements. The only requirement is *'large and identical amplification of both polarization channels'* separately using external phase modulations of single photon wave packets. Green squares are 50:50 polarizing beam splitters. Polarizing beam splitters #2, #3 are used to filter out any unwanted polarization states due to the presence of spontaneous emissions during amplification processes. These last two 'classical' beam splitters can ignore negligible quantum fluctuations entering through their unused ports while filtering these highly amplified beams in respective polarization channels. Thicker arrows indicate the amplified beams.

## 2. Description of the experimental framework of the proposed quantum cloning scheme.

### 2.1 General outlines of the operational structure

The schematic of this quantum mechanical cloning machine (QCM) is described in **Figure 1**. We assume that the incoming photon is linearly polarized such that the resultant polarization vector is in an unknown quantum superposition state as

$$|\boldsymbol{\Psi}\rangle = \alpha|\uparrow\rangle + \beta|\rightarrow\rangle \tag{1}$$

where $\alpha$ and $\beta$ are 'unknown' complex numbers such that $(\alpha)^2 + (\beta)^2 = 1$. The states $|\uparrow\rangle$ and $|\rightarrow\rangle$ are vertical and horizontal polarizations respectively and form an orthonormal, complete basis in Hilbert space of those two states. The net polarization of the state $|\boldsymbol{\Psi}\rangle$ is also at an unknown angle θ with respect to the vertical axis. In an actual experimental set up, one may also use a superposition of left and right circularly polarized light instead of $|\uparrow\rangle$ and $|\rightarrow\rangle$. Operationally speaking, one can design the QCM in the following manner as shown in **Figure 1**. Initially a polarizing beam splitter can separate the respective vertical $|\uparrow\rangle$ and horizontal $|\rightarrow\rangle$ polarizations of the incoming photon in two different channels, likely with some additional noise from the quantum vacuum entering through the unused ports of the beam splitter numbered '1'. Although, it was argued in the no cloning literature that there exists no QCM which can exactly amplify any arbitrary superposition of polarization states. However, it is really important to understand [1,7] that those same arguments do not actually rule out the possibilities of having

some devices which can separately amplify two independent, orthogonally polarized states $|\uparrow\rangle$ and $|\rightarrow\rangle$ light. Therefore, one can first – (a) split these orthogonal polarizations and (b) thereafter amplify the weak signals of single photon quantum states in $|\uparrow\rangle$ and $|\rightarrow\rangle$ separately. For example, one can then use phase sensitive amplification processes for each polarization states using degenerate parametric amplifier with a strong pump beam such that $\omega_{signal} = \omega_{idler} = \frac{\omega_{Pump}}{2} = \omega$. Thereafter, one can - (c) allow these two amplified beams to be incident on a cloning device 'D' (blue rectangle in **Figure 1**) without any additional phase lag introduced between the two polarization channels. The final stage (d) of this QCM is this device 'D' which can be made of any light sensitive excitonic material/structure having high quantum yields of linear optical absorption. In subsequent sections, we will now argue that if those two different sets of excitons, generated by two different light beams having orthogonal polarizations $|\uparrow\rangle$ and $|\rightarrow\rangle$, can be driven into two separate but interacting BECs, then one will be able[6] to create a quantum superposition of this two-component excitonic BEC along with some additional light-matter interactions to execute this quantum cloning operation.

It is to be strictly noted here that one cannot detect and/or track these photon(s) during the beam splitting, amplification and final mixing process in the excitonic device 'D'. As a result, one will not be destroying the quantum information embedded within vertical $|\uparrow\rangle$ and horizontal $|\rightarrow\rangle$ polarization states of the incoming photon by intentionally excluding these measurements induced direct disturbances to the quantum state. Such amplifications to really large numbers of photons, however, will be required to photo excite sufficiently large numbers of excitons which can produce two interacting excitonic BECs in this cloning device 'D' assuming that these excitons remain as composite bosons. The physical mechanisms of identical but separate and large amplifications of vertical $|\uparrow\rangle$ or horizontal $|\rightarrow\rangle$ polarization states in both channels to yield large photon numbers are important and will be discussed in subsections 2.2 and 2.3. Subsequent merging of these 'identically' amplified beams having $|\uparrow\rangle$ or $|\rightarrow\rangle$ polarization states in D to create an interacting BEC of a macroscopically large number of excitons and consequently generating the 'exact' photon replicas will be described in the next two subsections 2.4 and 2.5. Finally, we will discuss this experimental scheme in Section 4 and conclude with Section 5.

### 2.2 Identical but separate and large amplifications of vertical and horizontal polarization channels

As such, the number-phase uncertainty principle certainly complicates the precise experimental evaluation of quantum optical phase of a single photon before the amplification. Moreover, challenges [8-10] of essential degradation of signal-to-noise ratio during any amplification processes were well documented and subsequently a lot of in-depth studies were reported on this particular topic [11-15] in the last couple of decades. In fact, it is understood that amplification of any arbitrary superposition of both polarization states of a single photon is not possible using a deterministic, noiseless, and coherent operation. However, that restriction do not apply to isolated amplification of one single polarization state of light which will be used in this case. One may be tempted to try noiseless linear amplification (NLA)[11,16] using quantum scissors along with single photons of either $|\uparrow\rangle$ or $|\rightarrow\rangle$ to drive these NLA stages for the respective vertical $|\uparrow\rangle$ and horizontal $|\rightarrow\rangle$ polarization channels, as described in **Figure 1**a of Reference (11). However, the gains of NLAs will never be large enough to generate any substantially large number of photons to be able to photoexcite two different sets of macroscopically large number of excitons in order to produce two distinct set of BECs in the device 'D'.

As long as the physical amplification processes (represented as red rectangles in **Figure 1**) of vertical $|\uparrow\rangle$ and horizontal $|\rightarrow\rangle$ polarization states remain large and 'identical', the density of photo generated excitons in D, presumably with a 'known' electronic structure, can also remain proportional to these incident amplified intensities from both polarization channels. Such 'identical' amplification of both channels can be ensured by having a common and identical $|\eta|^2$ amplification factor during these 'intensity' amplification process. Besides, in our case, we would like to operate outside the domains of strong quantum optical fluctuations of both polarization channels of light which will further be used to create two separate, but interacting excitonic BECs [6]. This will be addressed in Section 2.5.

### 2.3 Pathways for amplification of single photon wave packet

A single photon is an ideal, amplitude squeezed number state. So, after the first polarizing beam splitter, one can think of using the single photon states $|\uparrow\rangle$ or $|\rightarrow\rangle$ in the two separate channels in **Figure 1** as a seeded signal for the phase sensitive, degenerate parametric amplification to generate squeezed states [10,17] as the output of the polarization selective amplification stages. Although, single photon Fock state does not have a well-defined phase in quadrature space. Nevertheless, phase lock can be maintained between the weak signal of a single photon with much more intense pump beam using the well-known quasi phase matching (QPM) [18] technique of non-linear optics. Here the momentum conservation necessary for phase-matching is adjusted using a periodically pooled structure. In our case, the relevant phase ($\Phi$) between the intense pump beam and weak input beam (single photon) can further be controlled during the amplification process by periodical phase modulation of the weak input signal from a single photon with a mirror on a piezoelectric [10,19] transducer stage - i.e. **Figure 1** of Reference (10). In reality, one can also use electro-optic modulator-based phase modulation [20] of single-photon wave packets along with the above-mentioned use of QPM to execute this. In fact, such external, periodic, phase modulation[10,19,20] of the weak input light signal from a single photon having $|\uparrow\rangle$ and $|\rightarrow\rangle$ polarizations in two separate channels can be the key step for relative phase locking between the strong pump beam and weak signal from the single photons along with quasi phase matching[18]. It will also be helpful in nullifying the effect of any unwelcome coupling to random, quantized vacuum modes of light entering through the first polarizing beam splitter marked as '1' in **Figure 1**. Moreover, such phase sensitive, degenerate parametric amplification [10] can always be pre-calibrated to have large, identical gains $|\eta|^2$ for both $|\uparrow\rangle$ and $|\rightarrow\rangle$ polarization channels. Unlike the phase insensitive amplification, in phase sensitive amplification [10] procedures, the gain (G) is always independent of the number of input photons even for few or single photons. Prior experimental calibrations can, therefore, be used to identically amplify one particular quadrature component of both polarization states $|\uparrow\rangle$ and $|\rightarrow\rangle$ at the cost of extra noise in the other quadrature. This calibration can also take in to account the transmission loss of photons in both channels from the first polarizing beam splitter #1 to D. Choice of respective polarization configurations (type 0, type I, type II etc.) of pump, signal and idler with respect to the non-linear crystal and the angle between the optic axis and propagation direction can be chosen during the actual experiment. The Gain (G) of such phase sensitive amplification [10] can increase exponentially $\left(\sim exp(\gamma z)\right)$ with propagation distance (z) for the amplified quadrature, where the gain coefficient $\gamma$ is proportional to non-linear coupling to the pump beam. For significantly high gains, a phase sensitive amplifier can also compensate for inevitable losses to approach a noise figure close to one [10] for the amplifying quadrature. Moreover, this gain (G) can be further enhanced using a laser like optical cavity as well. In this way, these phase sensitive, degenerate parametric amplifiers can generate 'noiseless' [10], large, identical amplifications separately, independently for both vertical $|\uparrow\rangle$ and horizontal $|\rightarrow\rangle$ polarization states in respective channels of **Figure 1**. Appropriate bandpass/dichroic filters can be used to remove the pump beam after amplifications. Spontaneous parametric emission is an obvious problem in case of phase insensitive amplifier for small photon numbers 'n'. However, in a phase sensitive amplification [10], the noisy quadrature components also decay exponentially with enhanced propagation distance. Additionally, any random noise from spontaneous emissions in such a parametric down-conversion process can always be filtered [21] out using two additional 50:50 polarizing beam splitters (#2 & #3 in **Figure 1**) before these are merged on excitonic material/structure D. Moreover, with strongly amplified beams in both the polarization channels, one can actually ignore the effects of vacuum field fluctuations entering through the open ports of these last two beam splitters (2 & 3) before generating separate spin selective excitons in D. In the next section, we will discuss how these two identically amplified $|\uparrow\rangle$ and $|\rightarrow\rangle$ states can be used to generate two different sets of excitons (electron-hole pairs) which can subsequently be tuned in a phase coherent way to form interacting, two-component BEC of excitons as prescribed in the past [6]. Besides, we are only interested in the all-important 'identical' but separate amplifications of both $|\uparrow\rangle$ and $|\rightarrow\rangle$ photons. So, it must be noted that the process of separate amplification of orthogonal polarization channels won't contradict the 'No-Go' theorem of having any deterministic single photon amplifier as those restrictions are applicable only and only for single photons in an arbitrary quantum superposition states like that given in Equation (1). In the hindsight, such identical but separate amplifications of individual polarization states $|\uparrow\rangle$ and $|\rightarrow\rangle$ of light are not in conflict with quantum mechanics at all.

### 2.4 Proportional generation of two sets of excitons and formation of a two-component BEC

Given that these identically but separately amplified photons of both polarization channels can be used to produce proportionate numbers of excitons or electron-hole pairs in D, this experimental proposal for cloning should be on track. Several crucial details of these experimental steps will now be discussed below. As mentioned above in Section (1), we assume that through linear optical absorption processes, each incoming photon having either vertical $|\uparrow\rangle$ and horizontal $|\rightarrow\rangle$ polarizations can produce only and only one exciton in the device D. We also assume that all or same fraction of these two different ensembles of excitons photo generated individually with vertical $|\uparrow\rangle$ and horizontal $|\rightarrow\rangle$ polarizations can be driven into two independent excitonic BECs having the quantum ground states as $|\psi_{\uparrow}\rangle$ and $|\psi_{\rightarrow}\rangle$ separately. Thereafter, another light source can be tuned in to generate weak 'Josephson' like coupling [6] between $|\psi_{\uparrow}\rangle$ and $|\psi_{\rightarrow}\rangle$ to form a two-component BEC as described in the next section. The material/structure of the cloning device 'D' must also be sensitive for such spin/polarization (even momentum) selective optical transitions as well as allow for coherent interactions between these two distinct excitonic BEC having quantum states of $|\psi_{\uparrow}\rangle$ and $|\psi_{\rightarrow}\rangle$. Finding an appropriate excitonic system within a suitable semiconductor material/heterostructure, however, can be really crucial for successful implementation of this experimental scheme.

### 2.5 Final steps of quantum cloning using bosonic clones of a two-component BEC

Any quantum phase fluctuations at the limit of large number of amplified photons or equivalently with large amplifications as $|\eta|^2 >> 1$ may also be ignored [22] for each polarization channels. In fact, ignoring [6,23] such quantum phase fluctuations of each polarization channels can be useful in developing a mean-field description of the two-component quantum ground state of excitonic matter wave. Following Reference (6), formation of these two-component excitonic BEC can be modelled so that the quantum fluctuation of 'excitonic' field operators can be ignored at the large exciton numbers. Likewise, the resultant single-particle ground state of this two-component interacting BEC can then be written as [6],

$$|\mathbf{\Psi}_1\rangle = \alpha'|\psi_{\uparrow}\rangle + \beta'|\psi_{\rightarrow}\rangle \tag{2}$$

where coefficients are $\alpha' = \eta\alpha$ and $\beta' = \eta\beta$ respectively, where $|\eta|^2$ is the common and identical multiplication factor for both polarization channels during the 'intensity' amplification process and one has $(\alpha')^2 + (\beta')^2 = |\eta|^2$ as the single-particle normalization factor. To repeat, here we are only assuming that $|\eta|^2$ can be kept identical for both polarization channels using proper experimental calibrations. This is necessary to ensure that the density of both sets of excitons (say $N_\uparrow$ and $N_\rightarrow$) in the quantum ground states of $|\psi_\uparrow\rangle$ and $|\psi_\rightarrow\rangle$ can also remain proportional to $|\alpha|^2$ and $|\beta|^2$ respectively such that $N_\uparrow = |\alpha'|^2 = |\eta\alpha|^2$ and $N_\rightarrow = |\beta'|^2 = |\eta\beta|^2$. It is assumed here that the number of excitons condensed in the isolated, independent ground states of $|\psi_\uparrow\rangle$ & $|\psi_\rightarrow\rangle$ are fixed and remain proportional to $|\alpha'|^2, |\beta'|^2$ respectively, so that one can define the order parameters for both $|\psi_\uparrow\rangle$ & $|\psi_\rightarrow\rangle$. As a result, the relative phase of vertical $|\uparrow\rangle$ and horizontal $|\rightarrow\rangle$ polarization channels can remain well defined. Such experimental calibrations are operationally possible with prior information of the nature of polarization states of the incoming photon as $|\uparrow\rangle$ and $|\rightarrow\rangle$ as the basis states of the input photon as well as with a suitable choice of the material/structure D. However, as mentioned above, these two different set of excitons generated using amplified photons having either vertical $|\uparrow\rangle$ and horizontal $|\rightarrow\rangle$ polarizations can further be tuned [6] with additional Josephson like 'weak' interactions to form a two-component BEC of a macroscopically large number (say) N excitons such that $N = N_\uparrow + N_\rightarrow$. Following that one can assume the quantum ground state of that two-component excitonic BEC can be given as shown below in Equation (3). This is similar to that described by Equation (7) of Reference (6) in the context of interacting BECs of cold atoms.

$$|\mathbf{\Psi}_{2-component-BEC}\rangle = |\mathbf{\Psi}_1\rangle \otimes |\mathbf{\Psi}_1\rangle \otimes \ldots |\mathbf{\Psi}_1\rangle = \frac{1}{\sqrt{N!}}[\alpha' a^\dagger + \beta' b^\dagger]^N |0\rangle \tag{3}$$

where $|0\rangle$ is the vacuum state of the two-component quantum superposition of excitonic BEC states and $a^\dagger, b^\dagger$ are creation operators of two distinct set of excitonic quantum ground state of BEC as $|\psi_\uparrow\rangle$ *and* $|\psi_\rightarrow\rangle$ respectively such that $(\alpha_N)^2 + (\beta_N)^2 = N|\eta|^2$ where $\alpha_N = \sqrt{N}\alpha', \beta_N = \sqrt{N}\beta'$ for respective set of excitons. In general, it was known that large N thermodynamic limit will always produce [23,24,25] this particular quantum state $|\mathbf{\Psi}_{2-component-BEC}\rangle$. So, this now completes the mapping of the initial quantum state of the single photon $|\mathbf{\Psi}\rangle$ to this formation two-component BEC of macroscopically large number of excitons using quantum statistical processes.

Detailed theoretical calculations to produce this final state of $|\mathbf{\Psi}_{2-component-BEC}\rangle$ were already reported[23,24,25] in a different context for condensation of Helium-3 as well. Therefore, here we refrain from writing down the same equations already elaborated in these past reports[6,23,24,25] as well as for sake of keeping a much wider generality in our quantum cloning procedure using such a two-component BEC at this stage. The relative phase fluctuations of the quantum field operators of $|\psi_{\uparrow}\rangle, |\psi_{\rightarrow}\rangle$ in the $|\mathbf{\Psi}_{\mathbf{2}-component-BEC}\rangle$ will decrease [23,24,25] as $N^{-\frac{1}{2}}$ and will be practically negligible at the large N limit of thermodynamic equilibrium. As a result, it will be possible[6,23,24,25] to produce this quantum state $|\mathbf{\Psi}_{\mathbf{2}-component-BEC}\rangle$ as an intermediate stage which can eventually be used generate multiple cloned copiesof the incoming photon with $|\mathbf{\Psi}\rangle$. This will be described at the end of this Section.

One may ask about the scenarios of going beyond the mean field approximations [6]. In that case, there can be further splitting of ground state energy levels [Equations (22) and (39) in Reference (6)] under certain approximations of the interaction parameters of the two distinct BEC ground state excitons $|\psi_{\uparrow}\rangle$ and $|\psi_{\rightarrow}\rangle$. It was also described there how the ground-state wave function can take the form of a Schrodinger-cat state under such circumstances. In a way, going beyond the domain of mean field approximation, one can even create [6] a macroscopic ''Schrodinger-cat'' state formed by two interacting Bose condensates. In principle, the bosonic cloning using the two-component BEC excitons will still be possible. However, exploring such situations are currently beyond the aim and scope of this study.

As mentioned above, we must also note that the linearity of optical absorption process in the excitonic cloning device 'D' (blue rectangle in **Figure 1**) is, however, really important. This will be required to prevent the generation of any higher order, multi-particle superposition such as $[|\psi_{\uparrow\uparrow}\rangle + |\psi_{\rightrightarrows}\rangle]$ etc., which can also form during the merging of amplified vertical $|\uparrow\rangle$ and horizontal $|\rightarrow\rangle$ light polarizations while generating two distinct sets of excitonic BECs with quantum ground states $|\psi_{\uparrow}\rangle, |\psi_{\rightarrow}\rangle$ respectively. This is because superposition states like $[|\psi_{\uparrow}\rangle + |\psi_{\rightrightarrows}\rangle]$ and $[|\psi_{\uparrow\uparrow}\rangle + |\psi_{\rightrightarrows}\rangle]$ etc., are qualitatively different from states $\frac{1}{\sqrt{N!}}[\alpha' a^{\dagger} + \beta' b^{\dagger}]^{N}|0\rangle$ which is required for the universal cloning operations using such two-component BEC [6,23,24,25].

Finally, we expect that some of these excitons in those 'identical', single-particle quantum state of $|\mathbf{\Psi}_{1}\rangle = \alpha'|\psi_{\uparrow}\rangle + \beta'|\psi_{\rightarrow}\rangle$ within that two-component BEC ground state having long range spatial and temporal coherence can radiatively decay in to photons having $|\uparrow\rangle$ and $|\rightarrow\rangle$ polarization states respectively with quantum efficiency $(\xi)$. This is because, photons emitted through the recombination of all these excitons in the 'identical' quantum ground state of this two-component, interacting BEC will have a narrow momentum space distribution as well as a narrow spectral line width. As a result, the emitted 'cloned' photons will be spontaneously phase coherent by themselves as consequence of this BEC. This can also happen via polarization conserving stimulated emissions. One can also use an additional optical cavity to trigger these stimulated emissions as well. Thereafter, we can finally recover the initial quantum superposition state $|\mathbf{\Psi}\rangle = \alpha|\uparrow\rangle + \beta|\rightarrow\rangle$ of

the incoming photon using the generic spatio-temporally coherent emissions from the quantum ground state of the two-component, interacting BECs of excitons as

$$|\boldsymbol{\Psi}_{\boldsymbol{Cloned}}\rangle = |\boldsymbol{\Psi}\rangle \otimes |\boldsymbol{\Psi}\rangle \otimes \ldots |\boldsymbol{\Psi}\rangle = \frac{1}{\sqrt{M!}}[\alpha'' c^{\dagger} + \beta'' d^{\dagger}\,]^{M}|0\rangle \tag{4}$$

where $c^{\dagger}, d^{\dagger}$ are creation operators of photons in $|\uparrow\rangle$ & $|\rightarrow\rangle$ states respectively and generated from the single-particle quantum state of the two-component BEC as defined in Equation (3). One can have $\alpha_M{}' = \sqrt{M}\,\alpha'' = \sqrt{M}\xi\alpha' = \sqrt{M}\,\xi\eta\alpha = C\alpha$ and $\beta_M{}' = \sqrt{M}\beta'' = \sqrt{M}\xi\beta' = \sqrt{M}\,\xi\eta\beta = C\beta$ such that $(\alpha_M{}')^2 + (\beta_M{}')^2 = |C|^2$ and $C = \sqrt{M}\,\xi\eta$, where $C$ can be the experimentally determined constant factor. As a result, within a normalization factor $C$ due to non-ideal quantum yields of the optical absorption and radiative emission at the cloning device 'D', one can, in principle, clone any incoming pure state in arbitrary quantum superposition like $|\boldsymbol{\Psi}\rangle = \alpha|\uparrow\rangle + \beta|\rightarrow\rangle$ as

$$|\boldsymbol{\Psi}_{\boldsymbol{Cloned}}\rangle \equiv |\boldsymbol{\Psi}\rangle \otimes |\boldsymbol{\Psi}\rangle \otimes .. |\boldsymbol{\Psi}\rangle \cong \frac{(\xi\eta)^M}{\sqrt{M!}}[\alpha c^{\dagger} + \beta d^{\dagger}\,]^{M}|0\rangle \tag{5}$$

Obviously, in practice, there will also be some Stokes shifts and the energy of the emitted photons can be red shifted from those of the absorbed one in the excitonic device 'D'. However, a prior knowledge of the photon energy and the choice of polarization basis of the incoming signal being used for communication and proper choice of a known/tailored electronic structure of the semiconductor material/device D can be helpful to match the energies of amplified photons suitably with respect to the incoming one. This energy matching process is not going to affect the overall fidelity of the quantum cloning scheme for the incoming arbitrary superposition state of a single photon. Here we are also assuming that the quantum ground state of this two-component exciton BEC in 'D' is not 'dark' or spin/momentum forbidden for light emissions due to selection rules. However, incident angles of amplified photon beams of two different polarizations can always be tuned [26] for a desired outcome.

## 3. Discussions

Above mentioned quantum copying scheme is an 'interdisciplinary' experimental proposal using quantum optics, light-matter interactions and quantum statistical effects in condensed matter physics of BEC of excitons or electron-hole pairs using a semiconductor material/structure schematically termed here as 'D'. This is possible because at low enough densities these excitons can be considered as bosons within a BEC and these will be exact quantum clones of each other in the first place. Most importantly, the overall procedures of amplification followed by photo generation of excitons (bosons) driven towards a macroscopically large two-component, interacting BEC state and subsequently their radiative recombinations are certainly not a part of any norm conserving unitary processes within a single Hilbert space. Therefore, in this proposed QCM, we are not directly violating the no-cloning theorem but just 'bypassing' it to make exact quantum clones of any arbitrary superposition of quantum states. In fact, the relative errors of generating the $|\boldsymbol{\psi_{2-component-BEC}}\rangle$ state as shown in Equation (3) will practically vanish at the large N thermodynamic limit. We are only using a series of quantum statistical mechanics-based light-matter interactions to generate the two-component, interacting excitonic BECs at the large N limit of thermodynamic equilibrium. This is possible once the individual polarization modes can be first separated, identically amplified and then eventually mixed to produce the quantum superposition state of two interacting BECs[6,23,24,25] of excitons which are tuned by Josephson like external light-matter interactions. We also argued how this experimental scheme of mapping an arbitrary quantum superposition on to the quantum ground state of a two-component BEC can prevent unwanted multi-particle superposition(s). We deliberated certain pathways for producing quantum cloning of photons first through generation excitons and then finally by radiative recombinations. We showed that these processes can be mediated by the quantum statistical nature of the quantum ground state of two-component BEC having macroscopically large number of phase coherent excitons. Formation of a superfluid BEC of macroscopically large number of excitons can resist[27] usual dissipation and decoherence associated with a multi-particle quantum system. Moreover, one can also extend this scheme beyond such photo generated excitonic ensembles to any other two-component BEC systems including exciton-polaritons, photons and even to atomic systems as well.

Most importantly, the expected experimental fidelity of such mapping can always be maintained by ensuring $\alpha_M{}' = \sqrt{M}\,\alpha'' = \sqrt{M}\xi\alpha' = \sqrt{M}\,\xi\eta\alpha = C\alpha$ and $\beta_M{}' = \sqrt{M}\beta'' = \sqrt{M}\xi\beta' = \sqrt{M}\,\xi\eta\beta = C\beta$ respectively with $C = \sqrt{M}\,\xi\eta$. This $C$ can be the experimentally tailored constant factor which can be kept identical in the QCM which involve the identically amplified input lights (i.e. for vertical $|\uparrow\rangle$ and horizontal $|\rightarrow\rangle$) fed to D, generation of corresponding excitons (i.e. $N_\uparrow$ in $|\psi_\uparrow\rangle$ & $N_\rightarrow$ in $|\psi_\rightarrow\rangle$ respectively) and then the cloned photon output as per Equation (5). As long as this $C$ remains *'identical'* for both polarization channels throughout the process, the density matrices $\rho_{input}$ for the input quantum state of Equation (1) and the $\rho_{output}$ for output quantum state of each single-particle components of the above Equation (5) are same $\left(\rho_{input} = \rho_{output} = \rho\right)$ within the above mentioned experimentally adjustable normalization factor described above. As a result, the expected fidelity of such quantum cloning in terms of can be given as

$$F\left(\rho_{input}, \rho_{output}\right) = \left[Tr\left(\sqrt{\sqrt{\rho_{input}}\rho_{output}\sqrt{\rho_{input}}}\right)\right]^2 = F(\rho, \rho) = 1. \quad (6)$$

Therefore, having such abundantly many bosonic clones of an interacting two-component BEC can increase the cloning fidelity and reduce errors for an exact quantum cloning to an insignificant level by producing a large number of clones.

In the hindsight, the security of quantum cryptography can still be recovered by 'operationally' moving from a simple 'qubit' like 2-level quantum system to a quantum system having large and unknown 'd' dimensional 'qudits' or even to an infinite dimensional continuum basis states to transfer information. This can make the process of identifying a suitable material/device system (as 'D' in **Figure 1**) to generate a well-defined d-component BEC mediated QCM enormously complicated, if not impossible for large, unknown and possibly a randomly variable 'd' for successive usages. Such qudit based quantum processors [28,29] and communication [30] devices are certainly being developed in the recent past.

## 4. Conclusions

Finally, formation of such a two-component, interacting BEC is not a norm conserving unitary evolution restricted within the domain of a single Hilbert space, but belongs to the domain of a collective [27] multi-particle, (N+1) dimensional Hilbert space. This is because the process of creating an interacting, two-component BEC excitons [31] through light-matter interactions belongs to the domain of quantum statistical mechanics. Thus generating bosonic clones using the quantum ground state of a two-component excitonic BEC is a suitable medium for experimental execution of 'exact' cloning of any arbitrary quantum state without having any direct conflict with basic premises of the no-cloning theorem[1,2] of quantum mechanics. To summarise again, this whole experimental proposal is centred primarily on (i) identical but separate and (ii) large amplifications of both vertical $|\uparrow\rangle$ and horizontal $|\rightarrow\rangle$ polarization channels separately to generate macroscopically large number of photons, then (iii) ignoring the phase fluctuations at large N thermodynamic limit, (iv) photo generation of a two-component[32] BEC of macroscopically large number excitons tuned with external but 'weak'[6] light-matter interactions within the semiconductor structure, finally (v) the radiative recombination of these phase coherent excitons from the BEC ground state leading to 'cloned' photons.

We also note that instead of pulsed [10], phase sensitive, degenerate parametric amplifier as proposed above, use of a continuous wave [33] pumped degenerate parametric amplifier may also be beneficial to work along with our experimental proposal. The large N thermodynamic limit can also offer many operational advantages – (a) firstly, it helps us to ignore phase fluctuations during the amplification of single photon, (b) secondly, it also facilitates in generating[6,23,24,25] the quantum ground state [Equations (3)] of two interacting BECs of macroscopically large N number of excitons in the mean field approximation as the ‘suitable’ mediator for the final quantum cloned state (Equation 5), (c) thirdly, such a quantum ground state [Equation 3] of a BEC of macroscopically large N number excitons can be represented as pseudo-orthogonal states on the Bloch sphere where bosonic enhancement of energy makes it amenable for faster[27] quantum gate operations at a timescale reduced by the factor N as well. Moreover, having a phase coherent BEC as the intermediate stage also helps in preventing the all-important quantum phase information from dephasing[6,23,24,25] within an experimentally realizable time scale.

Furthermore, exact cloning is also a sufficient condition for allowing communication of information between space-like separated points using entanglement. Therefore, by that same token, such possibilities of ‘exact’ quantum cloning of any maximally entangled state of photon, generation of which are ‘not’ limited [34,35] by the constraints of the linearity, unitarity and completeness of the 1st quantized wave functions within a single Hilbert space can, in principle, no longer prohibit superluminal [36,37], EPR[38] like signal communications using the above-mentioned scheme via the formation of a two-component BEC. However, this particular ‘interesting’ implications for fundamental physics alone should not be used as a pivotal argument against the proposed experimental scheme for quantum cloning described in this paper!

**ACKNOWLEDGMENTS**

SD acknowledges the Science and Engineering Board (SERB) of Department of Science and Technology (DST), India (Grants # DIA/2018/000029, CRG/2019/000412) for supports.

---